\documentclass[a4paper]{article}
\usepackage{authblk}
\usepackage{graphicx} 
\usepackage{amsmath}
\usepackage{siunitx}
\usepackage[labelformat=simple]{subcaption}

\usepackage{url}
\input{mymathphys.sty}
\input{myunits.sty}

\title{AMBER - A Strong-Interaction Facility at CERN\footnote{Submitted to Nuclear Physics News}}
\author[1]{B. Ketzer\thanks{Supported by BMFTR (Germany)}}
\author[2,3,4]{M. Chiosso\thanks{Supported by INFN (Italy)}}
\author[4]{the AMBER Collaboration}
\affil[1]{Universität Bonn, Helmholtz-Institut für Strahlen- und Kernphysik, 53115 Bonn, Germany}
\affil[2]{University of Torino, Dept. of Physics, 10125 Torino, Italy}
\affil[3]{Torino Section of INFN, 10125 Torino, Italy}
\affil[4]{CERN, 1211 Geneva 23, Switzerland}
\date{December 2025}

\begin{document}

\maketitle

\begin{abstract}
AMBER (NA66) is a fixed-target facility at the M2 beam line of CERN SPS, which performs worldwide unique research on the internal structure and the excitation spectrum of hadrons. 
The approved first phase of the experiment focuses on three main physics topics: (i) the measurement of the production cross section of antiprotons in $p-\text{He}$ and $p-p/d$ collisions over a wide energy range; (ii) the precise measurement of the electric form factor of protons at small momentum transfers using a high-energy muon beam; (iii) the determination of pion and kaon quark PDFs through Drell-Yan and charmonium production measurements with negative and positive meson beams.
Phase-2 will focus on measurements with an intense kaon beam.
The high-energy muon, pion and kaon beams required for these measurements are only available at CERN. 
\end{abstract}

\section{Introduction}
Despite the impressive success of the Standard Model, some of its key components remain surprisingly elusive. In Quantum Chromodynamics (QCD), the quantum field theory of the strong interaction, quarks and gluons interacting via the exchange of colour charge are the fundamental ingredients. Nevertheless, how exactly the complex properties of hadrons -- bound states of quarks and gluons -- emerge from these fundamental ingredients, remains one of the biggest unsolved problems within the Standard Model. One of the reasons for the difficulty is that we cannot simply disassemble a hadron into its constituents, as we could do for molecules, atoms, or even nuclei. The quarks and gluons remain confined inside hadrons; we cannot observe them as free particles outside of these. 

At the north end of CERN’s Super Proton Synchrotron accelerator complex (SPS), the experimental hall EHN2 hosts a general-purpose facility designed to address some of the long-standing questions in hadrons physics: the Apparatus for Meson and Baryon Experimental Research (AMBER). Building on the legacy of the COMPASS experiment, AMBER aims to shed light on key open issues in strong-interaction physics using high-intensity hadron and muon beams that are only available at the M2 beam line \cite{Adams:2018pwt,Adams:2676885}.

Approved by the CERN Research Board in December 2020, AMBER Phase-1 data taking started in 2023 and focuses on 
three flagship scientific goals: a precision measurement of antiproton-production cross sections relevant for indirect dark-matter searches; a new determination of the proton charge radius using elastic muon-–proton scattering; Drell–Yan and charmonium measurements to explore the inner structure of mesons and the emergence of hadron mass. 
After two years of data taking (2023, 2024) dedicated to the antiproton production measurements in proton--Helium ($p-\mathrm{He}$), proton--Hydrogen ($p-\mathrm{H}$) and proton--Deuterium ($p-\mathrm{D}$) collisions, the first integrated test run including every key element of the Proton Radius Measurement (PRM) setup has been successfully completed in late 2025, marking a major milestone on the way to the full PRM physics run in 2026. 
The break during the Long Shutdown 3 (LS 3) of the CERN accelerator complex will be used to prepare the additional detectors and installations required for the Drell-Yan measurement. 

Phase 2, currently foreseen to start in 2031, will focus on measurements with an intense kaon beam. This includes spectroscopy of strange mesons, measurements of the gluon structure of kaons and pions using meson-induced production of prompt photons, determination of the low-energy parameters of the kaon in ultra-soft collisions with nuclei, and measurements of meson charge radii in inverse kinematics. 
The increase in beam intensity and quality required for the measurements will be achieved as part of CERN's ongoing North Area Consolidation Program (NA-CONS), 
as well as further upgrades to the M2 beam line that have already been approved. 

\section{Antiproton production cross sections for dark matter search}
A substantial body of evidence suggests 
that the majority of matter in the universe is non-baryonic and electrically neutral. This component, known as dark matter, accounts for about $27\%$ of the total mass--energy budget of the universe. Its origin and nature remain among the most intriguing open puzzles in modern physics; a widely discussed hypothesis is that dark matter consists of weakly interacting massive particles (WIMPs), cold thermal relics from the Big Bang.
Indirect detection of dark matter candidates is based on the search for products of their annihilation or decay. These would appear as distortions in gamma-ray spectra and as anomalies in rare cosmic-ray components. In particular, antimatter components such as antiprotons, anti-deuterons and heavier(anti-)nuclei offer sensitivity to dark matter annihilation on top of the standard astrophysical production.

Interpreting galactic cosmic-ray data in search of potential exotic contributions from dark matter requires not only a realistic modelling of the sources and of the turbulence spectrum of the galactic magnetic field, but also reliable knowledge of the cross sections governing the production of antimatter components in interactions with the interstellar medium. Indeed, the dominant part of the antiprotons in the galaxy originates from inelastic scattering of cosmic-ray protons and helium nuclei off hydrogen and helium in the interstellar medium. 
Empirical modelling of these cross sections induces an uncertainty of about $15-20\%$ on the predicted antiproton flux. This contrasts with the unprecedented results at few percent level accuracy achieved by AMS-02 \cite{Aguilar:2016AntiprotonFlux}, on the International Space Station, in the GeV--TeV energy range.
In order to obtain a significant sensitivity to dark matter signals, accurate measurements of the \(\bar{p}\) production cross section in \(p-p\) and 
\(p - {}^{4}\mathrm{He}\) collisions over a wide energy range from 10~GeV to a few~TeV
are thus of fundamental importance.

In 2023, AMBER carried out an extensive measurement of the antiproton production cross section using a proton beam incident on a liquid helium (${}^{4}\mathrm{He}$) target. This provided new data for the reaction $p+{}^{4}\mathrm{He} \rightarrow \bar{p} + X$, covering an unprecedented beam-momentum range from 60 to 250 GeV/c, crucial for modelling the cosmic antiproton flux \cite{Adams:2018pwt,Adams:2676885}.
In 2024, the programme was extended to include \(p-p\) and \(p-\mathrm{D}\) collisions at beam momenta of 80, 160 and 250 GeV/c. Using liquid hydrogen and deuterium targets, these measurements are specifically designed to investigate potential isospin-asymmetry effects in the production of antiprotons versus antineutrons.
The experimental setup for these measurements is largely based on the COMPASS spectrometer, with the Ring Imaging Cherenkov Detector (RICH-1) used for antiproton identification \cite{Tessarotto:2014C09011}. 

\begin{figure}[p]
    \centering
    \includegraphics[width=1.\textwidth]{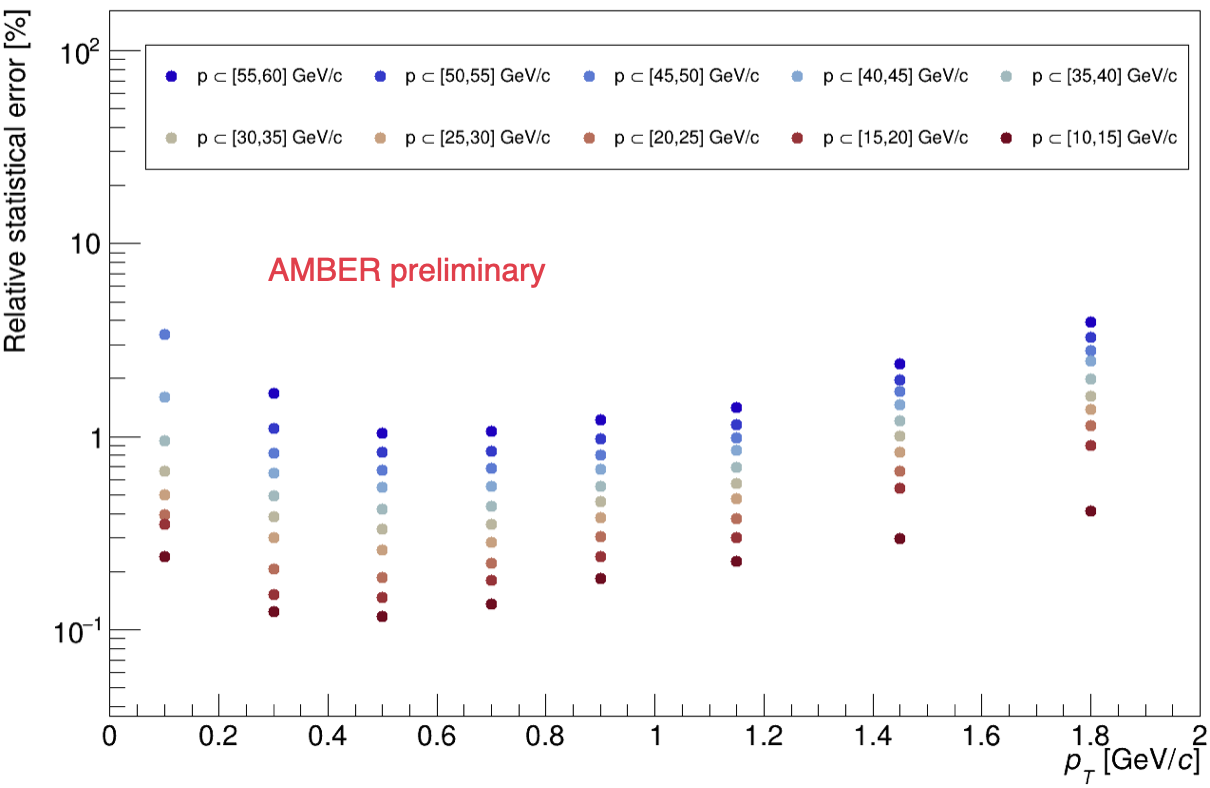}
    \caption{Statistical uncertainties on the number of antiprotons in different bins in momentum and $p_T$ for the 2023 $p-\mathrm{He}$ data collected at $\sqrt{s_{\mathrm{NN}}} = 18.9$\,GeV/c.  
}
    \label{fig:expStat}
\end{figure}

In Fig.~\ref{fig:expStat} the statistical uncertainties on the number of antiprotons in different bins in momentum and $p_T$ are shown for the 2023 $p-^4\mathrm{He}$ data collected at $\sqrt{s_{\mathrm{NN}}} = 18.9$\,GeV/c.
The phase-space coverage is excellent with statistical errors $ < 1\% $ in most kinematical bins. The data analysis is ongoing and first results will be published in 2026.

\section{Measuring the proton radius}
The spatial distributions of the electric charge inside hadrons are a fundamental property, obviously related to the confining force between quarks. 
As was shown in the pioneering measurements of Robert Hofstadter et al. \cite{Hofstadter:1955ae}, the mean squared charge radius $\left\langle r_E^2\right\rangle$ can be determined from the measurement of the electric form factor in elastic scattering of leptons on protons, and by extrapolating it to zero 4-momentum transfer squared $Q^2$. The textbook wisdom of a large radius of $\left\langle r_E^2\right\rangle^{1/2}=0.88\,\textrm{fm}$, determined with increasing precision over decades by electron-proton elastic scattering \cite{A1:2010nsl}, was called into question in 2010 by a high-precision measurement of the Lamb Shift in muonic hydrogen \cite{Pohl:2010zza} that resulted in a small radius of  $0.84\,\textrm{fm}$, with an uncertainty much smaller than the one from electron scattering. This unexpected result was the origin of the so-called proton radius puzzle, which triggered widespread discussions on systematic uncertainties, theoretical corrections, or even new physics. 

\begin{figure}[p]
    \centering
    \begin{subfigure}[t]{\textwidth}
        \includegraphics[width=\textwidth]{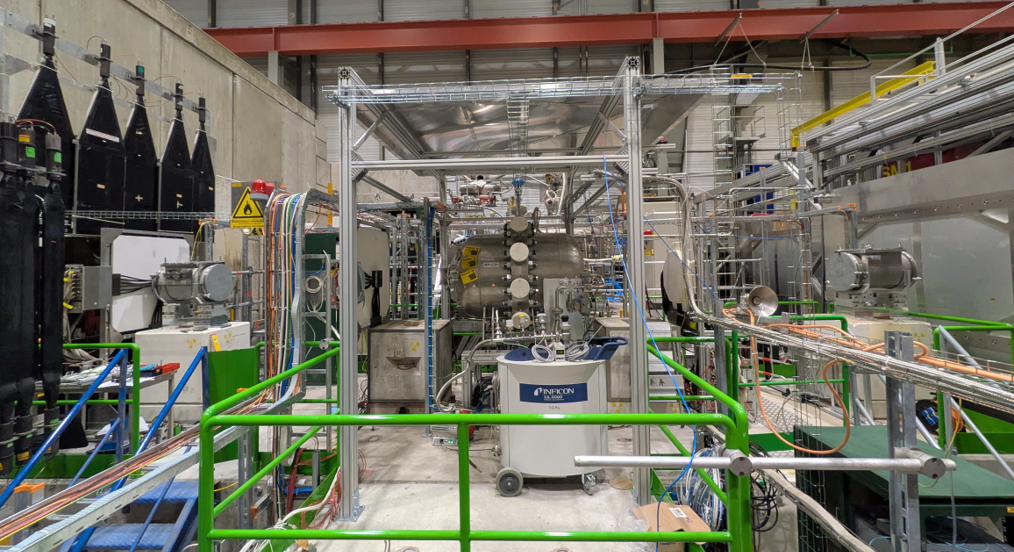}
        \caption{}
        \label{fig:AMBER_PRM_target_region}
    \end{subfigure}
    \begin{subfigure}[t]{\textwidth}    
        \includegraphics[width=1.\textwidth]{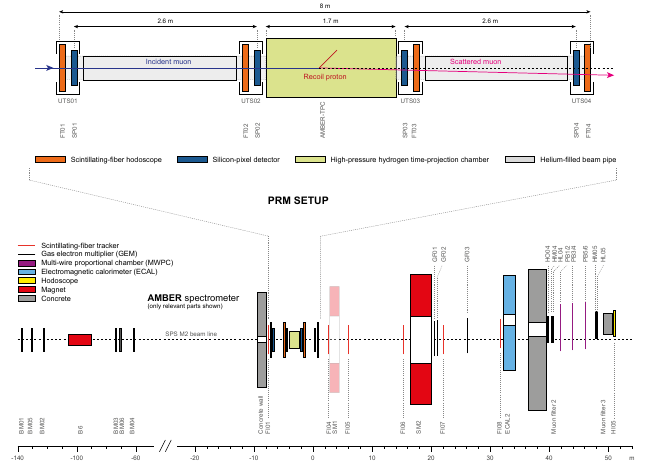}
        \caption{}  
        \label{fig:prm:AMBER_PRM_Setup} 
    \end{subfigure}
    \caption{AMBER experimental setup for the PRM programme. (a) Photograph of the target region. (b) (Top) Schematics of the target region with the high-pressure TPC in the centre. (Bottom) Complete spectrometer.}
\end{figure}

AMBER set out to provide a new and independent precision measurement of this quantity by using high-energy muon elastic scattering on protons rather than low-energy electrons or muons \cite{Adams:2018pwt,Adams:2676885}. Such a high-energy muon beam ($100\,\GeV$) is available exclusively at CERN SPS. 
Using muons instead of electrons is highly advantageous, as several experimental systematic effects and also theoretical (radiative) corrections are considerably smaller.
The measurement requires the installation of a set of completely new detectors in the target region of AMBER, as shown in Fig.~\ref{fig:prm:AMBER_PRM_Setup}.

\begin{figure}[p]
    \centering
        \begin{subfigure}[t]{0.45\textwidth}
        \includegraphics[width=\textwidth]{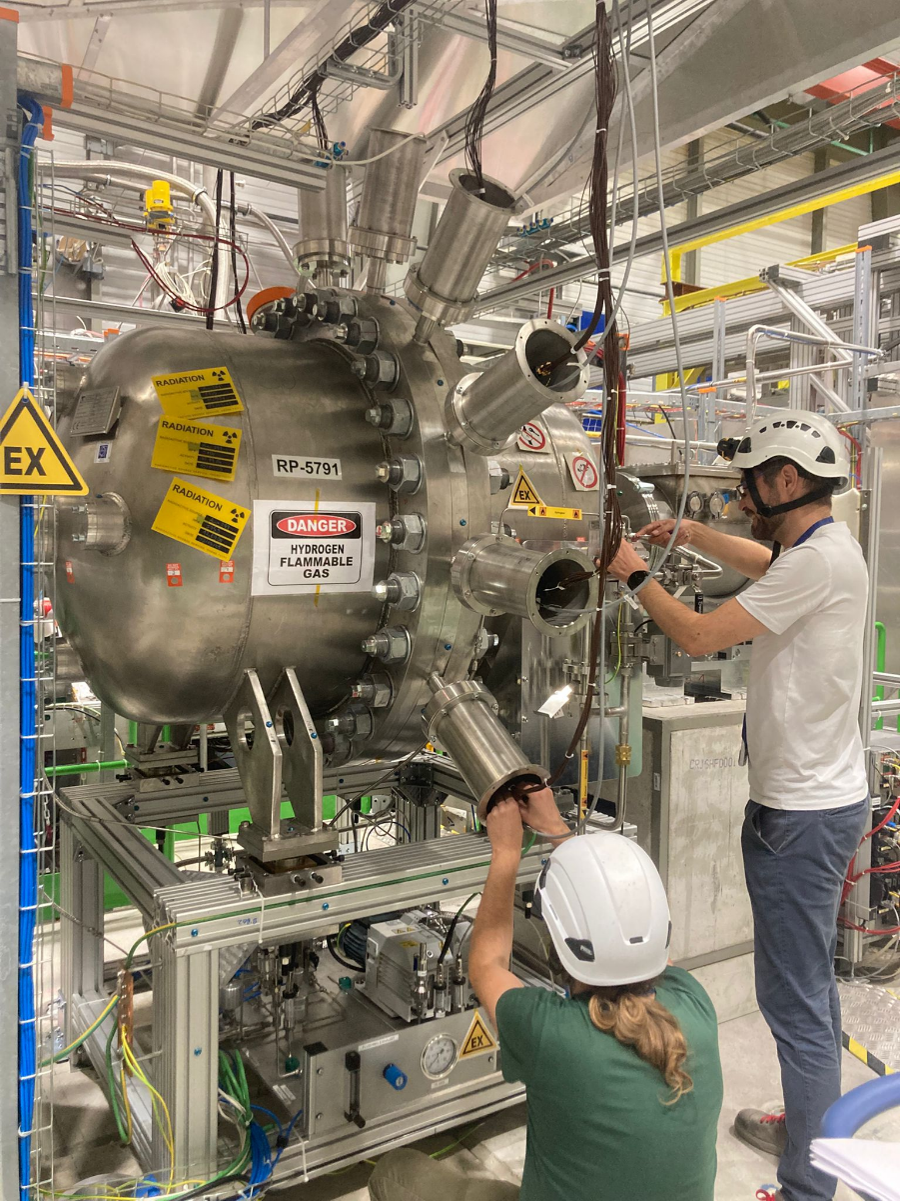}
        \caption{}
        \label{fig:TPC_installation}
    \end{subfigure}
    \begin{subfigure}[t]{0.54\textwidth}       
        \includegraphics[trim={0.0cm 10.0cm 0.0cm 11.0cm}, clip,width=\linewidth]{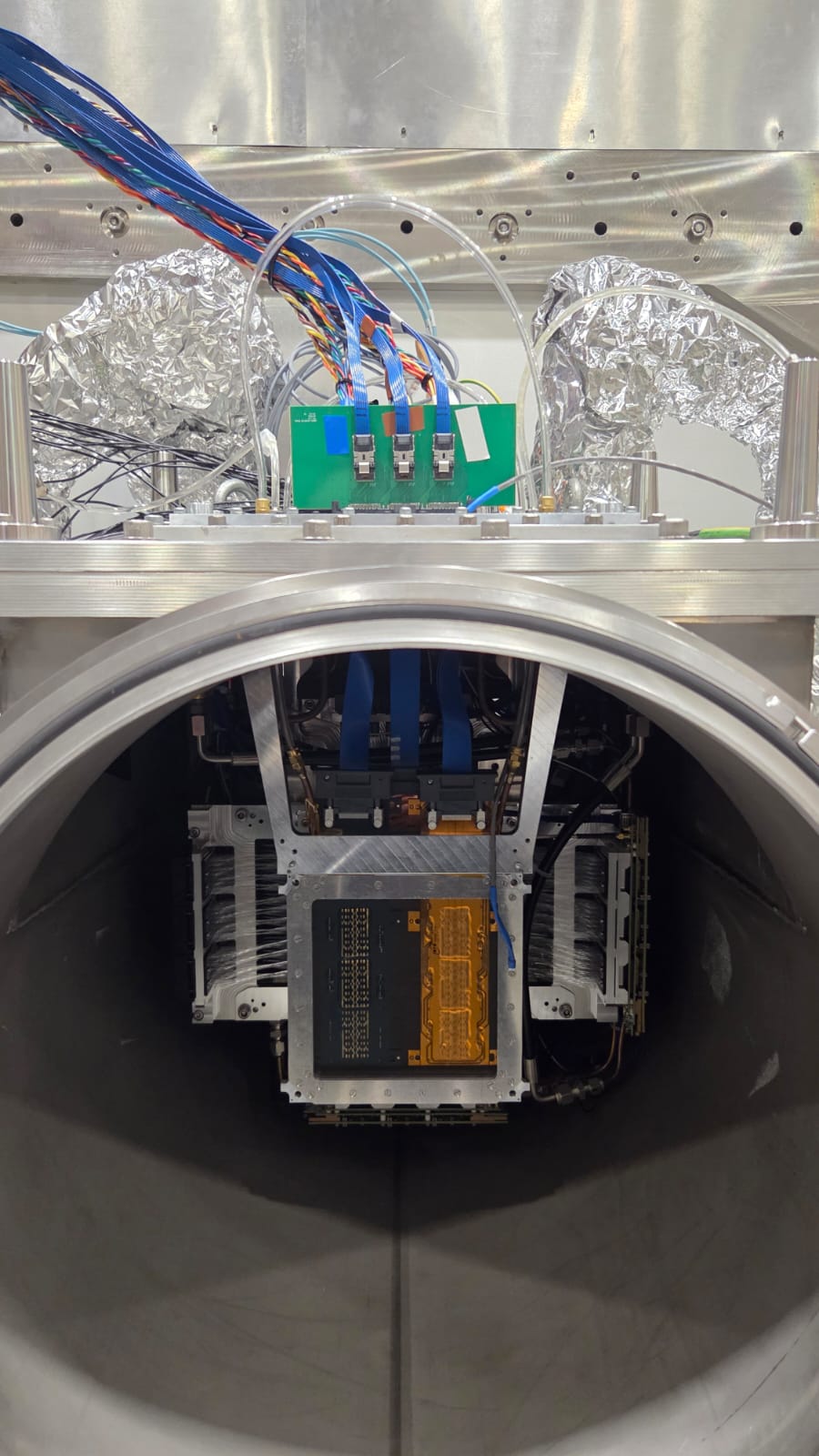}
        \caption{}
        \label{fig:UTS_in_beam}
    \end{subfigure}   
    \caption{(a) Final installation work on the AMBER TPC in the experimental hall, (b) inner view of a Unified Tracking Station (UTS): an SPD plane in the foreground, mounted in front of an SFH station in the background. \label{fig:AMBER_PRM_Target_Region}}
\end{figure}

The new setup features cutting-edge silicon-pixel and scintillating-fibre technology for charged-particle tracking and an active-target using ultrahigh-purity hydrogen gas at a pressure of $20\,\text{bar}$. 
The required low momentum transfer between the beam muon and the target proton means that the recoil energies of the proton can be as low as $500\,\text{keV}$. In order to detect the recoil protons, the target medium is used simultaneously as detector: a Time Projection Chamber (TPC) filled with pure hydrogen. The ionization charge produced by the recoil proton drifts in an electric field of $100\,\text{kV/cm}$ over a maximum distance of $40\,\text{cm}$ towards a segmented anode plane. No internal charge multiplication is applied in order not to spoil the energy resolution of the instrument: the TPC operates in ionization mode, with the induction gap being separated from the drift region by a Frisch grid. 
To maximize the luminosity, two TPC cells 
are installed in a common high-pressure vessel, with the readout anodes of both cells attached to a central flange, as can be seen in Fig.~\ref{fig:TPC_installation}. 

While the energy of the recoil proton, together with the energy of the incoming beam muon, is in principle sufficient to measure the differential cross section for elastic scattering, the clean selection of elastic-scattering events requires the identification of the scattered muon and the precise determination of the scattering vertex. The first task is taken over by the AMBER spectrometer featuring a dipole magnet, charged-particle tracking using GEM and MWPC detectors, muon identification, and an electromagnetic calorimeter for the detection of radiative events. The latter task is performed by a brand-new vertexing system made of four so-called Unified Tracking Stations (UTS), mounted up- and downstream of the TPC, see Fig.~\ref{fig:prm:AMBER_PRM_Setup} (top). Each station comprises three planes of Monolithic Active Pixel Sensors (MAPS) of the ALPIDE type with $8\,\mu\text{m}$ spatial resolution (Silicon Pixel Detector, SPD), and four planes ($2\times X$, $2\times Y$) of $0.5\,\text{mm}$ thick scintillating fibres (Scintillating Fibre Hodoscope, SFH), individually read out by Silicon Photomultipliers, to resolve hit association ambiguities in the $10\,\mu\text{s}$ long time window of the pixel detector. 
Figure~\ref{fig:UTS_in_beam} shows a view inside a UTS.

In order to cope with the vastly different time responses of the various detectors and to allow for intelligent event selection, the detector front-end electronics and the data acquisition (DAQ) system was moved from a triggered to a free-streaming system. From the continuous data stream, an online high-level trigger logic selects slices corresponding to detector-specific time windows that are written to disk. From these data, physics events are extracted using a 4-dimensional tracking algorithm. 

During the 2025 run, all the subsystems anticipated for the 2026 physics run were successfully operated in the full beam environment. 
In 2026, AMBER plans to perform the first $\mu$–p elastic-scattering measurement with the complete setup, and thus to add an important new piece to the proton radius puzzle. 

\section{Investigating the quark structure of mesons}
In the past 25 years, the COMPASS experiment made fundamental contributions to the quark and gluon spin structure of nucleons by deep-inelastic scattering of muons off polarized proton and deuterium targets, and through Drell–Yan measurements with a pion beam on a transversely polarised proton target. The internal structure of pions and kaons, the Goldstone bosons of dynamical chiral symmetry breaking, is far less known, and relies on very limited experimental data.
AMBER will address this knowledge gap by using high-intensity pion and kaon beams interacting with nuclear targets to study the quark structure of mesons through the Drell–Yan process, in which an antiquark or quark in the meson annihilates with a quark or antiquark in the nucleus, producing a lepton-pair in the final state.

The unique feature of the M2 beam line is that it provides both positive and negative meson beams with high energies up to $200\,\text{GeV}$. This allows us to separate valence and sea quark contributions. For the measurement, the target region of AMBER will be equipped with three $^{12}\text{C}$ targets of $25\,\text{cm}$ length, interspersed with a Silicon vertex detector, and a hadron absorber downstream of the target. The di-muon pair is detected and reconstructed in the upgraded charged-particle spectrometer. Particle identification in the beam at rates up to several hundred MHz is an essential requirement of the measurement. To this end, the readout capabilities of the beam Cherenkov detectors (CEDARs) are presently being improved.
The AMBER measurement represents the most comprehensive study of meson structure planned for this decade. 

\section{Conclusions and outlook}
AMBER is a new QCD facility at CERN's Super Proton Synchrotron that performs globally unique measurements related to the emergence of hadronic and nuclear degrees of freedom in bound systems of quarks and gluons. Its key features are the use of high-intensity muon and hadron beams of both charge signs hitting an experiment-specific target region, and a high-resolution, high-acceptance spectrometer for the detection of charged and neutral final-state particles. A free streaming DAQ in combination with an online high-level triggering system allows for the selection of complex event topologies based on different detector types. The AMBER physics program covers a wide range of topics ranging from lowest $Q^2$ measurements as the determination of the proton electric form factor to high-$Q^2$ reactions to study the internal quark and gluon structure of unstable particles such as the pion and the kaon. The results are expected to lead to a significant improvement in the understanding of QCD as the theory of strong interactions. 
 
\bibliographystyle{utphys}
\bibliography{biblio,amber}
\end{document}